\documentclass[aps,epsf]{revtex4}
\def\Vect#1{\mbox{\boldmath $#1$}}
\usepackage{graphicx}
\usepackage{amsmath}
\begin{document}

\draft
\title{Rheology of a granular gas under a plane shear}
\author{Kuniyasu Saitoh\footnote{Present address: DeNA,
Sasazuka, Sibuya, Tokyo, 151-0073, JAPAN} 
and Hisao Hayakawa\footnote{The corresponding
author. e-mail: hisao@yukawa.kyoto-u.ac.jp , Address after July1, 2006,
Yukawa Institute for Theoretical Physics, Oiwake-cho, Kitashirakawa,
Sakyo, Kyoto 606-8502, JAPAN.}
}
\address{Department of Physics at Yoshida-South Campus, Kyoto University,  Yoshida-Nihonmatsu, Sakyo,
Kyoto 606-8501, JAPAN}
\begin{abstract}
The rheology of a two-dimensional granular gas under a plane shear is
 investigated.
From the comparison among the discrete element method, the simulation of
a set of hydrodynamic equation, and the analytic solution of the steady
 equation of the hydrodynamic equations,
it is confirmed that the fluid equations derived from the kinetic theory
 give us 
accurate results even in relatively high density cases.

\end{abstract}
\maketitle

\section{Introduction}

Granular materials consist of macroscopic dissipative
particles. In some cases the granular material behaves as an unusual fluid.
Although to understand the rheology of the granular fluid is practically
important,  our understanding on the rheology
is not enough. There are several
reasons to have poor understanding on the rheology
of granular flows: (i) The separation of the length scale between particles
and the fluid motion  is not enough, (ii) there are some cases that
the fluid region coexists with the solid-like region, and (iii) most of
experiments are strongly 
affected by boundary conditions and the
external field.
Nevertheless, it is believed that 
rapid granular flows for relatively dilute granular gases can be
described by a set of hydrodynamic equations at Navier-Stokes order whose
transport coefficients can be calculated by the kinetic theory.\cite{goldhirsch}

To maintain a granular gas we need to add an external field. The
simplest steady situation of the granular fluid is achieved by the balance
between an external shear and inelastic collisions between
particles. This system is appropriate to investigate what the
constitutive equation for
the granular fluid is. The kinetic theory may suggest that 
the stress-strain relation 
can be described by that at Navier-Stokes order,
though the transport coefficients can be functions of the position.\cite{poeschel,brilliantov04,brey98}

About 50 years ago, Bagnold\cite{Bagnold} suggested that the granular fluid is
characterized by $\tau\propto \dot{\gamma}^2$ where $\tau$ and
$\dot\gamma$ are the shear stress and the shear rate (the strain),
respectively. This stress-stain relation is known as
Bagnold's scaling  and is different from the conventional Newtonian
relation which is $\tau\propto \dot{\gamma}$.
 Recently, Pouliquen\cite{Pouliquen} and Silbert {\it et al.}\cite{silbert} have reconfirmed the quantitative
relevancy of Bagnold's scaling in granular flows on inclined slopes. 
Mitarai and Nakanishi\cite{Mitarai2} have demonstrated that the kinetic
theory can be compatible with Bagnold's scaling, when they assume that
the the temperature is a slaving variable of the velocity and the
density.  Santos {\it et al.}\cite{santos04} also indicate that
Bagnold's scaling is valid for steady dilute granular gases without the
influence of the gravity in the uniform shear flow (USF), though  the
transport coefficients such as the viscosity and the heat conductivity
are different from those in homogeneous cooling state\cite{brey98}.
However, we still do not know whether Bagnold's scaling is relevant in other
situations. 

We should recall that it is difficult to keep granular
gases in experimentally relevant situations because of the existence of
gravity. 
Recently, to remove the effects of 
gravity, the rheology of dense granular flows under the plane shear with a
constant pressure has been studied and a new  scaling has been
reported\cite{GDR,cruz}. These  studies are important but particles
are not in a gas state, {\it i.e.} each particle is in contact with many
other particles simultaneously.  The analysis of such process is challenging
but we do not have any good tool to analyze it at present. Here, we
focus on the granular shear flows without multi-body contacts
 in a constant volume
container to discuss quantitative relevancy of the kinetic theory.   

The purpose of this paper is to know the relevancy of the kinetic theory
for a  granular gas with moderate density under the plane shear
in a constant volume container to characterize the rheology of granular
fluids. 
For this purpose, we investigate whether (i)  the kinetic theory
is 
relevant for the relatively dense granular gas,  and (ii) 
the tangential contact and the rotation of particles are irrelevant
except for the boundary layers.
It should be
noted that Bagnold's scaling is no more relevant for flows with 
non-uniform shear
rate in the steady granular flow. In fact, to satisfy $\tau\propto 
\dot\gamma^2$, the shear rate 
$\dot\gamma$ should be uniform, because the shear stress and the
pressure are spatially uniform in the steady state.   To investigate the above problem, we use the discrete
element method (DEM) for particles' simulation in a plane shear problem
of dense granular gases in which particles are confined in a constant
volume box (Section II). We adopt the
constitutive equation of hydrodynamics 
derived by Jenkins and Richman\cite{JR} for
non-rotational particles in Section III. 
In Section IV we solve a set of hydrodynamic equations with the above
constitutive equation numerically and compare the result with the
result of DEM. We also check the renormalization theory of the
restitution coefficient developed by Yoon and Jenkins\cite{Yoon}.  In
Section V we obtain the analytic solution of the steady hydrodynamic
equations to verify the quantitative relevancy of the constitutive
equation. In section VI we discuss our result and the relevancy of
Bagnold's scaling. 
In Section VII, we conclude our results. In
Appendix, we briefly explain the method to determine the tangential
restitution constant as a function of the incident angle. 

\section{DEM simulation}

\subsection{DEM model}

The discrete element method (DEM) is one of the standard methods to
simulate granular motions.\cite{cundall}  DEM is applicable to most of situations of granular dynamics even when particles are almost
motionless and contact with many other particles. 
We adopt DEM to simulate a granular fluid to check (i) the validity of
kinetic theory based on the reliable model, and (ii) the effects of rotation of
particles for the granular fluid. 

In this paper, we focus on a two-dimensional motion of granular particles
under a plane shear. 
We adopt the linear spring model for the repulsion 
with the normal stiffness $k_n$ and the tangential stiffness $k_t$,
and the normal and the tangential viscous coefficients $\eta_n$ and
$\eta_t$, respectively.

  Let us consider a colliding pair of 
two disks $i$ and $j$ of the diameter $\sigma$  
  and the mass $m$ at the position $\Vect{x}_k$ with the velocity
  $\Vect{c}_k\equiv
\dot{\Vect{x}}_k$  and the angular velocity $\Vect{\omega}_k$ for
  $k=i$ or $j$. If the particles are in contact, the overlap distance
\begin{equation}
\Delta_{ij}\equiv 2\sigma-|\Vect{x}_i-\Vect{x}_j|
\end{equation}
must be positive. The relative velocity at the contact point is
\begin{equation}
\Vect{c}_{ij}={\Vect{c}_i}-{\Vect{c}_j}+
\frac{\sigma}{2}\Vect{n}_{ij}\times (\Vect{\omega}_i+\Vect{\omega}_j)
\end{equation}
with 
the normal unit vector $\Vect{n}_{ij}\equiv
(\Vect{x}_i-\Vect{x}_j)/|\Vect{x}_i-\Vect{x}_j|$.
Introducing the normal velocity
$c_n^{ij}=\Vect{n}_{ij}\cdot\Vect{c}_{ij}$, the tangential velocity
$c_t^{ij}=\Vect{t}_{ij}\cdot\Vect{c}_{ij}$, the tangential displacement
$w_t^{ij}=\int_{t_0}^t ds c_t^{ij}(s)$ with the tangential unit vector
$\Vect{t}_{ij}$ satisfying $\Vect{t}_{ij}\cdot\Vect{n}_{ij}=0$, the
normal and the tangential forces $F_n^{ij}$ and $F_t^{ij}$ are
respectively given by
\begin{eqnarray}
F_n^{ij}&=& mk_n\Delta_{ij}-m\eta_n v_n^{ij} \quad {\rm for}{~} \Delta_{ij}>0 ,  \\  
F_t^{ij}&=& min(h_t,\mu |F_n^{ij}|)sign(h_t^{ij}) ,
\end{eqnarray}
where 
$h_t^{ij}\equiv-m k_t w_t^{ij}-m\eta_t c_t^{ij}$ with Coulomb
friction constant $\mu$, $min(a,b)$ is the function to select the smaller
one between $a$ and $b$, and $sign(x)=1$ for $x>0$ and $sign(x)=-1$ for
$x<0$.  The total repulsive force at the contact  can be represented as $\Vect{F}_{ij}=F_n^{ij}\Vect{n}_{ij}+F_t^{ij}\Vect{t}_{ij}$.

Thus, the equation of motion of particle $i$ is described by
\begin{eqnarray}\label{2-3}
	m\dot{\Vect{c}}_i&=&\sum_{j\ne i}\Vect{F}_{ij} ,\\
	I\dot{\Vect{\omega}}_i&=&\displaystyle\frac{\sigma}{2}\sum_{j\ne i}
\Vect{n}_{ij}\times\Vect{t}_{ij}F_t^{ij} ,
\label{2-4}\end{eqnarray}
where 
$I=m\sigma^2/8$
is the moment of inertia.
We integrate (\ref{2-3}) and (\ref{2-4}) in terms of the second order
Adams-Bashforth with the time interval $\delta t=4.0\times 10^{-4}
 (2\sigma/U)$.

Through the paper we adopt the following parameters as 
$k_n=3.0\times 10^3 (U/\sigma)^2$, $k_t=k_n/4$, $\eta_n=3.0 (U/\sigma)$, $\eta_t=\eta_n/2$ and
$\mu=0.20$, where $U$ is the shear speed at the boundary. 
These parameters lead to the normal restitution
constant $\bar{e}=0.85$ and the tangential restitution 
$\beta\simeq -1+1.12442\cot\gamma$ for 
$\gamma\le \gamma_c$ and  $\beta=\beta_0\simeq 0.769235$ for $\gamma>\gamma_c$ where $\gamma$ is 
the incident angle of two colliding disks and the critical angle
$\gamma_c$ is given by 
$\cot \gamma_c\simeq 1.56734$ (see eq.(\ref{3-27}) in Appendix).
As shown in Appendix, the tangential restitution constant $\beta$
can be approximated by\cite{walton} 
\begin{equation}\label{walton}  
\beta \simeq
\left\{
 \begin{array}{ll}
  -1+\mu (1+\bar{e}) \cot\gamma \left(1+\frac{m\sigma^2}{4I}\right)
&  (\gamma \ge \gamma_{c})\\
   \beta_{0} &  (\gamma \le \gamma_{c}).
 \end{array}
\right.
\end{equation}
We also note that the realistic value of Coulomb friction constant
in both disks and spheres is  $\mu \le 0.2$\cite{stronge,rosato,kuninaka}
Thus, the renormalization theory of the restitution constant may be
applicable to many realistic situations.

\subsection{Set-up}

\begin{figure}[h]
\centering
\includegraphics[width=10cm]{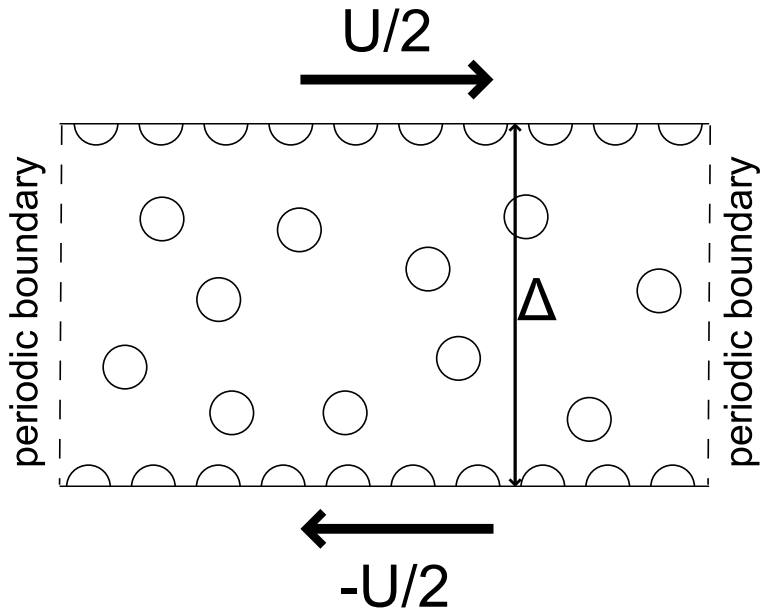}
\caption{The configuration of our simulation.}
\label{fig:4-1}
\end{figure}

Simulations of granular particles under the plane shear have been
performed by many researchers, but many of them\cite{Campbell,Dave,Tan}
assume Lees-Edwards boundary condition\cite{Lees} which may not be
adequate to consider the effects of physical boundary.  On the other
hand, Babic\cite{babic97b}, Popken and Cleary\cite{Popken} have simulated 
sheared granular flows 
confined in frictional flat boundaries, but their simulations are
restricted to the cases for small systems and almost elastic particles. 
Kim\cite{kim95} has indicated that the density of particles near the
boundary is higher than that in the bulk region for his simulation of a
small system with the flat frictional boundary, while particles are
accumulated in the center region for a larger system. 
Louge\cite{Louge94} has simulated
a three dimensional shear flow on the flat frictional boundary to examine
the boundary condition proposed by Jenkins\cite{jenkins92}, but Louge is 
mainly interested in the behavior of flux, the stress ratio as the
functions of volume fraction and the restitution constant. Recent
papers by Xu {\it et al.} for an experiment\cite{Xu1} and 
a simulation\cite{Xu2} examine the validity of three dimensional kinetic
theory by Jenkins and Richman\cite{JR3} under asymmetric shears in
the presence of a streamwise body force, where  they
obtain reasonable agreements between the theory and the observations in
both the experiment and the simulation.

As long as our knowledge, we do not know papers to discuss the validity
of kinetic theory in a transient dynamics and a symmetric shear without
a streamwise body force with an enough large system.  
Thus, we adopt the following setup of our DEM simulation shown in
Fig.\ref{fig:4-1}.  The
system is confined in a two-dimensional container. Without including the 
effects of the air and the gravity 
we add a symmetric shear with the shear speed $U$
as shown in Fig.\ref{fig:4-1}. The parameters are fixed as the number of 
particles $N=5000$,  the linear dimension of the system in $y$ direction 
$\Delta=180\sigma$ and the mean
area fraction $\bar\nu=0.121$.
The boundary condition of $x-$direction is periodic.
We non-dimensionalize all quantities  by the diameter $\sigma$ for the length scale, $m$ for the
mass, and the inverse of the shear rate $2\sigma/U$ for the time scale.

We introduce some fixed particles on the wall to reproduce the bumpy
boundary. The reason why we adopt the bumpy boundary is to
avoid the large amount of slips on the wall under a physical situation.  
In our simulation we start from an initial condition without the shear.  
Then, the wall at $y=\Delta/2$ obeys the equation of motion in $x$ direction
\begin{equation}
M_w\frac{d\Vect{c}_W}{dt}=-m\gamma_w (\Vect{c}_W-U\Vect{e}_x/2)+\Vect{F}_{ex} ,
\end{equation}
where $\Vect{c}_W$ and $\Vect{e}_x$ are respectively the actual wall velocity
and the unit vector along $x-$direction. $M_w$, $\gamma_w$ and
$\Vect{F}_{ex}$ are the mass of wall $M_w=5.0\times 10^6 m$, the
relaxation rate $\gamma_w=10 U/(2\sigma)$, and 
the force acting on the wall by the collision  between mobile particles
and the wall, respectively.

\subsection{Simulation}
\begin{figure}[htb]
\includegraphics[width=12cm]{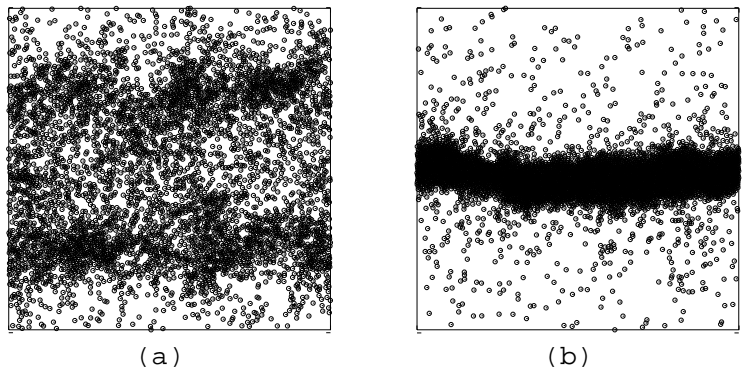}
\caption{The time evolution of the particles' configurations from (a) to 
 (b) for $\bar{\nu}=0.121$. }
\label{fig:4-2}
\end{figure}

The initial condition is prepared as that the configuration of particles 
 is at random  and the velocity distribution function
 obeys Maxwellian. Figure~\ref{fig:4-2} is the time evolution of
 particles' configuration for $\bar\nu=0.121$. Two shallow clusters
 appear near the wall, and move  to the center region of the
 container. Then, the two clusters merge to form a  big cluster. A similar
 behavior can be observed in the simulation under the Lees-Edwards boundary
condition.\cite{Tan} 

\begin{figure}[htb]
\includegraphics[width=10cm]{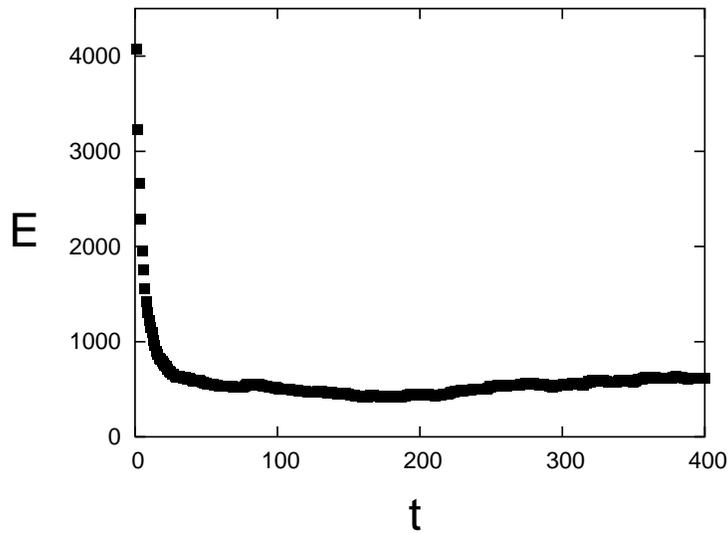}
\caption{The time evolution of the kinetic energy, where the units of
 time and the energy are $2\sigma/U$ and $m U^2/2$, respectively.
}
\label{fig:4-3}
\end{figure}

The time evolution of the total kinetic energy $E$ in a typical situation is shown
in Fig.\ref{fig:4-3},  where the energy is defined by
\begin{equation}
E(t)=\frac{1}{2}\sum_i (m c_i^2+I\omega_i^2) .
\end{equation}
In this figure, the initial energy is larger than the steady value, but
we have checked that the qualitatively similar results can be obtained
even when we start from the smaller energy about $E(0)=1200$ in the
dimensionless unit.
It is characteristics that the total kinetic energy is relaxed to be almost
a constant value quickly, but there is the slow  
evolution of hydrodynamic fields.

The hydrodynamic variables 
are the local area fraction $\nu(\Vect{r},t)\equiv\pi\sigma^2n(\Vect{r},t)/4$ 
with the number density $n(\Vect{r},t)$, the  velocity field 
$\Vect{v}(\Vect{r},t)$ and the granular temperature $T(\Vect{r},t)$ which is defined by
\begin{equation}
T(\Vect{r},t)=\frac{1}{2n}\int d\Vect{c} (\Vect{c}-\Vect{v})^2f(\Vect{r},\Vect{c},t) .
\end{equation}
In our simulation we divide the system into square cells with the linear 
dimension $\sigma$. Then, we can check to what cell  each particle belongs.
Thus, the measured  area fraction in our simulation is given by 
$\nu(\Vect{r},t)\equiv \sum_{i\in C} \pi \sigma^2/A$ where the summation
is taken over the center of particle $i$ existing  in the cell $C$ at $\Vect{r}$ with
the  area $A=\sigma^2$.
Similarly, the velocity field $\Vect{v}(\Vect{r},t)\equiv \sum_{i\in C} \Vect{c}_i$ is
the local average of the velocity of  particles. The temparture
field is also calculated by
\begin{equation}
T(\Vect{r},t)=\frac{1}{2n}\sum_{i\in C} (\Vect{c}_i-\Vect{v})^2.
\end{equation}



\begin{figure}[htb]
\includegraphics[width=16cm]{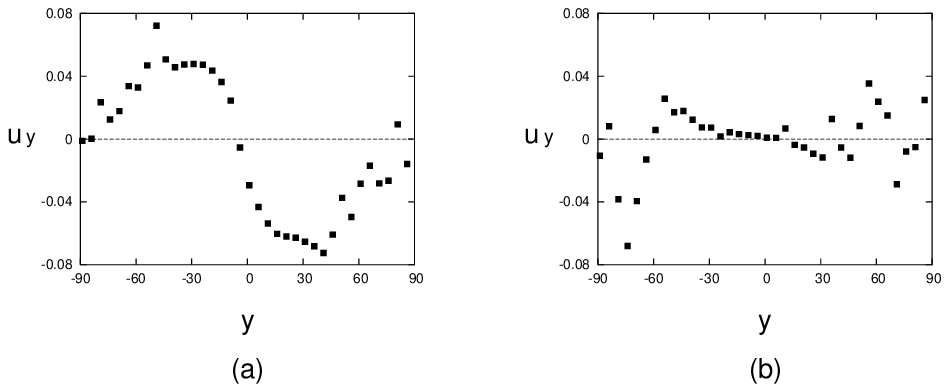}
\caption{The time evolution of  dimensionless 
$u_y$ with (a) at $t=60$ and (b)  
 at $t=380$, where the units of the time and $u_y$ are $(2\sigma/U)$ and 
 $U/2$, respectively.}
\label{fig:4-8}
\end{figure}


The time evolution of hydrodynamic variables can be summarized as
follows. Corresponding to Fig.2, the local area fraction becomes large near the boundary at the initial stage, and the shallow
clusters move to the central region of the container with growing the peak
density. Finally, the two clusters merge  to form a compact
cluster at the center $y=0$.

The interesting behavior can be observed in the velocity field and
the granular temperature. The $x$-component of the velocity field is almost zero between two clusters during
 the time evolution, and such plug region is narrower as the distance of
 two clusters is closer. Even in the steady state, the velocity gradient
 in the central region is smaller than that near the boundary.
The behavior of $y$-component of the dimensionless velocity field $u_y$ is also characteristic,
because this quantity shows the tendency to move to the central region
and is relaxed to be zero in the steady state (Fig.\ref{fig:4-8}).
Similarly, the temperature field in the central region is smaller than
that near the boundary. However, the minimum value of temperature
decreases with time and reaches almost zero in the central region in the
steady stage. 
The results of our DEM simulation except for $u_y=2v_y/U$
will be shown in Sections IV and V
through the comparison of DEM with hydrodynamic simulations or the
theoretical results in the steady state.

The result of our simulation may give us suspicious impression of the
validity of kinetic theory for this system, because (i) the density 
in the cluster is near the closest packing $\pi/2\sqrt{3}\simeq 0.907$ 
in the steady state, (ii) the
particles are almost motionless in the clusters or between
clusters. Our result is contrast to the result by Xu {\it et
al.}\cite{Xu1,Xu2} where there is no definite motionless clusters because 
of the existence of a streamwise body force. In
their case, the mean density is much higher than that in our case, and
the temperature is a nearly constant.  
Nevertheless, as will be shown, our system can be described
by the hydrodynamic equation derived from the kinetic theory.

\section{Hydrodynamic equations}

The purpose of this paper is to verify the validity of the granular hydrodynamic equation 
derived from the kinetic theory. Although there are two standard
methods, Chapman-Enskog method \cite{Chapman,brilliantov04,brey98} and
Grad expansions \cite{Grad}, most of established results
 are limited to low density cases in the derivation of hydrodynamic equations.
However, as shown in the previous section, we have to adopt the kinetic
theory for dense granular gases.  The Chapman-Enskog method by Graz\'{o} and
Dufty\cite{GD} and Lutsko\cite{lutsko05} predicts transport coefficients in dense granular gases.
On the other
hand, Jenkins and Richman derive hydrodynamic equations based
on Grad expansion and give 
transport coefficients\cite{JR,JR3}. Although the treatment by Jenkins
and Richman\cite{JR} does not take into account the contraction of the phase
space volume in each collision, the theory is suitable for our purpose because it gives us explicit expressions of the
transport coefficients in the two-dimensional  dense granular gases.

Jenkins and Richman\cite{JR}, and Lun\cite{Lun1} derive sets of
hydrodynamic equations which include the angular velocity, the spin
temperature as well as the density, the translational velocity and the
granular temperature.  The equations include the couple stress and the
collisional loss of spin energy. These sets of hydrodynamic equations are
categorized as the micropolar fluid mechanics which was originally
proposed by Cosserat and Cosserat for the description of the elastic
materials\cite{Cosserat}. Application of the concept of micropolar fluid mechanics to atomic gases  are developed by Dahler
and his coworkers\cite{Dahler}. The micropolar fluid mechanics is
applied to granular flows by Kanatani\cite{kanatani}, Lun\cite{Lun1},
Babic\cite{Babic}, Hayakawa\cite{haya2000} 
and Mitarai {\it et al}.\cite{Mitarai3}. 
The importance of the excitation of the spin on the boundary is 
indicated by Jenkins\cite{jenkins92}, but the effects of the spin can be
decoupled with the translational velocity in the bulk region. In
particular, Babic\cite{Babic} indicated that the coupled stress induced by the
collisions between circular particles is
canceled. Therefore, we believe that the effect of spins
can be absorbed in the  boundary condition and the
renormalized restitution constant.

Recently, Jenkins and Zhang\cite{JZ}
have proposed a scheme of the renormalized  restitution constant as the
result of the absorption of the rotation of
particles. Yoon and Jenkins\cite{Yoon} have extend the scheme to
two-dimensional flows as
\begin{equation}\label{e_eff}
e \simeq \bar{e}-\mu+2\mu^2(1+\bar{e}) .
\end{equation}
The validity of three dimensional theory\cite{JZ} has been tested by Xu
{\it et al.}\cite{Xu2} and Jenkins and Zhang\cite{JZ}. 
The latter  is
consistent with Lun and Bent\cite{LB} in part.
However, the quantitative validity of Yoon and Jenkins\cite{Yoon} has
not been confirmed yet.
In addition, some recent papers suggest that the spin effects are
relevant in granular flows.
For instance, Goldhirsch {\it et al.}\cite{Goldhirsch2} have indicated that
the equipartition between spin energy and the translational energy is
violated in a recent paper, and 
Gefen and Alam\cite{Gefen_Alam} discuss the linear stability of sheared
micropolar fluid.  Therefore, we need to judge whether the concept of
micropolar fluid is necessary for the description of granular fluids. 

In this paper, we adopt the renormalization procedure of the restitution
constant proposed by Yoon and Jenkins\cite{Yoon} to verify the validity
of their scheme. 
Thus, the restitution constant $e$ appears in
hydrodynamic equations is different from $\bar{e}$ of DEM, where the relation 
between two restitution constants is given in eq.(\ref{e_eff}). Namely,
 $\bar{e}=0.85$ in DEM corresponds to $e=0.798$ for $\mu=0.20$ 
in hydrodynamic equations.

The advantage to adopt the renormalization is that
hydrodynamic equations can be simplified as
\begin{eqnarray}
  D_t \rho &=& -\rho \Vect{\nabla}\cdot\Vect{v}  ,\\
  \rho D_t \Vect{v} &=& -\Vect{\nabla}\cdot\Vect{P} ,\\
  \rho D_t T &=& -\Vect{P}: (\Vect{\nabla}\Vect{v})-\Vect{\nabla}\cdot\Vect{q}
 -\chi ,
\end{eqnarray}
where $\rho=nm$ is the mass density and $D_t=\partial_t+\Vect{v}\cdot\nabla$.
Here $(i,j)$ component $P_{ij}$ of the pressure tensor $\Vect{P}$ is
expressed as the function of the bulk viscosity $\xi$ and the shear
viscosity $\eta$
\begin{equation}
P_{ij}=[p-\xi(\Vect{\nabla}\cdot\Vect{v})]
\delta_{ij}-\eta\hat{D}_{ij}
\end{equation}
at the Navier-Stokes order, 
where $\delta_{ij}=1$ for $i=j$ and $0$ for otherwise, 
$\hat{D}_{ij}=(\nabla_i v_j+\nabla_j v_i)/2$. Here $\Vect{q}$ represents the heat flux
which can be expanded as
\begin{equation}
\Vect{q}=-\kappa\Vect{\nabla}T-\lambda\Vect{\nabla}\rho ,
\end{equation}
where $\kappa$ is the heat conductivity and the transport coefficient
$\lambda$ disappear at $e=1$. The collisional loss rate of the energy $\chi$
can be represented by
\begin{equation}
\chi=\frac{1-e^2}{4\sigma\rho_p\sqrt{\pi}}\rho^2
 g(\nu)T^{1/2}[8T-3\sqrt{\pi}\sigma T^{1/2}(\Vect{\nabla}\cdot\Vect{v})] ,
\end{equation}
where $\rho_p=4m/(\pi \sigma^2)$ is the mass density of a particle.

Let us non-dimensionalize the time, the position, the velocity and the temperature as
\begin{equation}
 t=\frac{2\sigma}{U}t^*, \hspace*{2em}
 \Vect{x}=\sigma\Vect{x}^*, \hspace*{2em}
 \Vect{v}=\frac{U}{2}\Vect{u}, \hspace*{2em}
 T=\frac{U^2}{8}\theta 
\end{equation}
Thus, the non-dimensional pressure tensor, the heat flux  and the collisional loss rate of energy 
are respectively given by
\begin{equation}
 P_{ij}=\frac{\rho_p U^2}{4} P^*_{ij}, \quad
\Vect{q}=\frac{\rho_p U^3}{8} \Vect{q}^*, \quad
\chi =\frac{\rho_p U^3}{8\sigma}\chi^* .
\end{equation}
Here the dimensionless quantities are written as
\begin{eqnarray}
P^*_{ij}&=&[p(\nu)\theta-\xi(\nu)\theta^{1/2}(\Vect{\nabla}^*\cdot\Vect{u})]
\delta_{ij}-\eta(\nu)\theta^{1/2}\hat{D}^*_{ij} , \\
\Vect{q}^*&=&-\kappa(\nu)\theta^{1/2}
\Vect{\nabla}^*\theta-\lambda(\nu)\theta^{3/2}\Vect{\nabla}^*\nu , \\
 \chi^*&=&\frac{1-e^2}{4\sqrt{2\pi}}\nu^2 g(\nu)\theta^{1/2}[4\theta-3\sqrt{\frac
  {\pi}{2}}\theta^{1/2}(\Vect{\nabla}^*\cdot\Vect{u})] .
\end{eqnarray}
The explicit expressions of $p(\nu)$, $\xi(\nu)$, $\eta(\nu)$, $\kappa(\nu)$ and $\lambda(\nu)$ 
obtained by Jenkins and Richman\cite{JR} 
are summarized in table~\ref{Subfunction} with the radial distribution function\cite{volfson} 
\begin{equation}\label{CS-f}
g(\nu)=g_c(\nu)+\frac{g_f(\nu)-g_c(\nu)}{1+\exp[-(\nu-\nu_0)/m_0]},
\end{equation}
where $g_c(\nu)=(1-7\nu/16)/(1-\nu)^2$ and
$g_f(\nu)=[(1+e)\nu(\sqrt{\nu_c/\nu}-1)]^{-1}$ with $\nu_c=0.82$,
$\nu_0=0.7006$ and $m_0=0.0111$.
The choice of $g(\nu)$ is not be unique. For example, we expect that a similar
result can be obtained by using the radial distribution function in
ref.\cite{torquato}. 
Thus, the dimensionless hydrodynamic equations are reduced to
\begin{eqnarray}
\label{hydro1}
  D_t \nu &=& -\nu\Vect{\nabla}\cdot\Vect{u} ,\\
  \nu D_t \Vect{u} &=& -\Vect{\nabla}\cdot\Vect{P} ,\\
  \displaystyle \frac{1}{2}\nu D_t \theta &=& -P_{ij}(\nabla_{i}u_{j})-\Vect{\nabla}\cdot\Vect{q}-\chi .
\label{hydro3}
\end{eqnarray}
From here, the mark $*$ to represent dimensionless quantities is eliminated.

\begin{table}\label{table5.1}
\caption[]{The dimensionless transport coefficient by Jenkins and Richman}
\label{Subfunction}
 \begin{center}
  \begin{tabular}{rcl}
\toprule
	&&\\
   $p(\nu)$ &=& $\displaystyle \frac{1}{2}\nu[1+(1+e)\nu g(\nu)]$ \\
   &&\\
   $\xi(\nu)$ &=& $\displaystyle \frac{1}{\sqrt{2\pi}}(1+e)\nu^2 g(\nu)$ \\
   &&\\
   $\eta(\nu)$ &=&
   $\displaystyle \sqrt{\frac{\pi}{2}}[\frac{1}{7-3e}g(\nu)^{-1}+\frac{(1+e)(3e+1)}{4(7-3e)}\nu+(\frac{(1+e)(3e-1)}{8(7-3e)}+\frac{1}{\pi})(1+e)\nu^2 
   g(\nu)]$ \\
   &&\\
   $\kappa(\nu)$ &=&
   $\displaystyle \sqrt{2\pi}[\frac{1}{(1+e)(19-15e)}g(\nu)^{-1}+\frac{3(2e^2+e+1)}{8(19-15e)}\nu+(\frac{9(1+e)(2e-1)}{32(19-15e)}+\frac{1}{2\pi})(1+e)\nu^2
   g(\nu)]$ \\
   &&\\
 $\lambda(\nu)$ &=& $\displaystyle -\sqrt{\frac{\pi}{2}}\frac{3e(1-e)}{16(19-15e)}[4g(\nu)^{-1}+3(1+e)\nu]\frac{1}{\nu}\frac{d(\nu^2 g(\nu))}{d\nu}$\\
	&&\\ 
\hline
  \end{tabular}
 \end{center}
\end{table}

\section{Simulation of hydrodynamic equations}

\subsection{The outline of our simulation}

To verify the accuracy of a set of hydrodynamic equations
(\ref{hydro1})-(\ref{hydro3}) derived from
the kinetic theory by Jenkins and Richman\cite{JR}, we simulate
hydrodynamic equations. Since the separation of particles' scale and the
hydrodynamic scale is not enough, each grid in two-dimensional space
cannot contain enough number of particles to define hydrodynamic
variables. Therefore, we focus on the field equations which
have translational symmetry in
$x$-direction. Thus, all quantities only depend on $y$ and
$t$. However, we should note that we keep $x$-component of velocity. The
second purpose of the simulation of hydrodynamic equations is to obtain
a reduced set of equations  to recover the qualitative accurate results
to describe the metastable dynamics after the total energy is relaxed to
a constant value. 

The method of the discretization of continuous variables is based on the
standard procedure.  We adopt the classical Runge-Kutta scheme for the time derivative with
$\delta t=0.01$ 
and the second order 
accuracy of the spatial derivative of a hydrodynamic variable $\Psi$ as 
\begin{equation}
	\frac{\partial \Psi}{\partial y}=\frac{\Psi_{j+1}-\Psi_{j-1}}{2h} ,
\quad
	\frac{\partial^2 \Psi}{\partial y^2}=\frac{\Psi_{j+1}-2\Psi_{j}+\Psi_{j-1}}{h^2} ,
\end{equation}
where $h$ is the grid displacement with $h/\Delta=1/180$
 and $y=j h$ with $j=0,\pm 1, \pm 2, \cdots$.
 It should be noted that we do not
 have to solve Poisson equation for the pressure because the fluid is
 compressible and the pressure is completely determined by the equation
 of state.

\subsection{The boundary condition}

We adopt the boundary condition proposed by Johnson and
Jackson\cite{Boundary2}. We define the slip velocity on the boundary as
$\Vect{u}_{sl}=\Vect{u}-\Vect{u}_w$, where $\Vect{u}_w=\pm \Vect{e}_x$ at
$y=\pm \Delta/(2\sigma)$. Let $\Vect{t}$ and $\Vect{n}$ be
the tangential unit vector and the normal unit vector to the wall,
respectively.  Thus,
the conservation of the linear momentum on the wall is given by 
\begin{equation}
\label{8.4}
	-\Vect{n}\cdot\Vect{P}\cdot\Vect{t}=\frac{\pi}{4}\phi\Omega(\nu,\theta)|\Vect{u}_{sl}| ,
\end{equation}
where 
$\pi/4$ is originated from $m=\pi \rho_p\sigma^2/4$.
Here, $\phi$ is the roughness parameter and $\Omega(\nu,\theta)$ is the collisional frequency
between the wall and the particles. The expression $\Omega$ is assumed to be
\begin{equation}
\Omega(\nu,\theta)=\nu g(\nu)\theta^{1/2} ,
\end{equation}
where the prefactor is absorbed in the roughness parameter.
On the other hand, the energy balance on the wall can be expressed by
\begin{equation}
\label{8.6}
	\Vect{n}\cdot\Vect{q}=-\Vect{u}_{sl}\cdot\Vect{P}\cdot\Vect{n}-\Gamma(\nu,\theta)
\end{equation}
where  $\Gamma(\nu,\theta)$ is the energy loss rate in terms of the inelastic collisions
between particles and the wall, which may be represented as
\begin{equation}\label{8.7}
\Gamma(\nu,\theta)=\frac{\pi}{4}\Phi \Omega(\nu,\theta)\theta
	=\frac{\pi}{4}\Phi\nu g(\nu) \theta^{3/2} ,
\end{equation}
where $\Phi$ is the hardness parameter of the wall.
In our simulation we adopt $\phi=0.20$ and $\Phi=0.24$ as fitting parameters. 

The reason why we adopt the boundary condition by Johnson and
Jackson\cite{Boundary2} is that their condition is simple. The more
precise treatment for the boundary condition can be seen in
ref.\cite{jenkins01}. When we adopt Jenkins' boundary condition, the
number of fitting parameters may be reduced.

When we represent these boundary conditions as
\begin{equation}
F_{b1}(\Psi, \partial_y\Psi)=0
\end{equation}
and the formal solution of this discrete equation can be formally solved as
\begin{equation}
\Psi_N=F_{b2}(\Psi_{N-1},\Psi_{N-2}) ,
\end{equation}
where $N=\Delta/2\sigma$ is the grid number on the boundary with
symbolic functions $F_{b1}$ and $F_{b2}$. 
From the consideration of the symmetry in $y$-direction, we have
 $\nu(y)=\nu(-y)$, $\theta(y)=\theta(-y)$,
and $\Vect{u}(y)=-\Vect{u}(-y)$. Thus, it is enough to discuss the
boundary condition at $y=\Delta/2$.
From (\ref{8.4}) we may obtain the $x-$component of the velocity field
\begin{equation}
u_{x;N}=\displaystyle \frac{u_{x;N-2}+4\phi h \nu_{N-1} g(\nu_{N-1})/\eta(\nu_{N-1})}
			{1+4\phi h \nu_{N-1} g(\nu_{N-1})\eta(\nu_{N-1})} ,
\end{equation}
where the area fraction $\nu_N$ on the boundary is assumed to be
\begin{equation}
\nu_N=2\nu_{N-1}+\nu_{N-2}
\end{equation}
to suppress the gradient of the density field.
Similarly, eq.(\ref{8.6}) becomes
\begin{equation}
\theta_{N}=\displaystyle \frac{\theta_{N-1}\lambda(\nu_{N-1})(\nu_{N-2}-\nu_{N})}{\kappa(\nu_{N-1})}
+2h\frac{[\phi (1-u_{x;N})^2-\phi\theta_{N-1}]\nu_{N-1}g(\nu_{N-1})}{\kappa(\nu_{N-1})} .
\end{equation}
It is obvious that $y$ component of the velocity field satisfies 
\begin{equation}
u_{y;N}=0 .
\end{equation}

\subsection{The result of numerical simulation for the complete set of hydrodynamic equations}

\begin{figure}
\centering
\includegraphics[width=15cm]{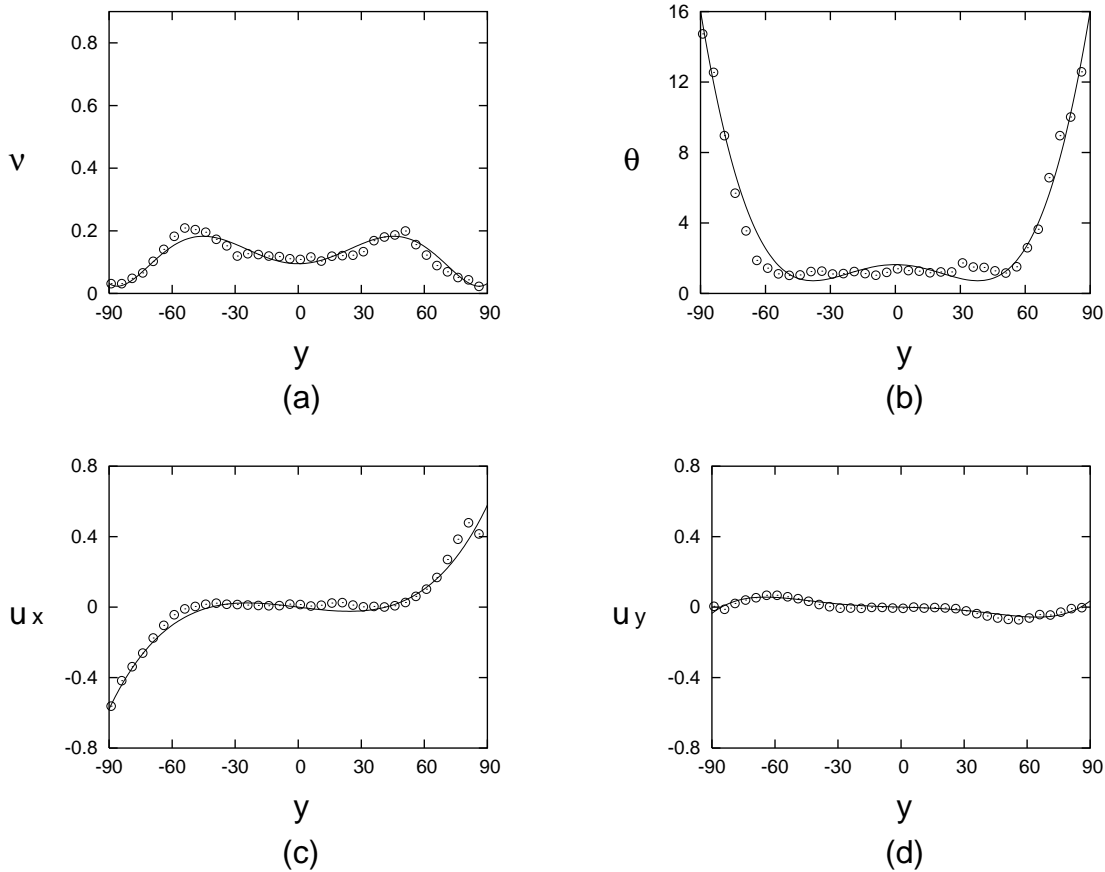}
\caption{The initial conditions of hydrodynamic equations (solid lines)
 and the corresponding data of DEM (open circles) at $t=20$. The
 solid lines for $\nu$ and $\theta$ are the polynomials of even powers
 of $y$ until $y^6$, while the lines for $u_x$ and $u_y$ are the
 polynomials of the odd powers of $y$ until $y^5$.}
\label{fig:8-1}
\end{figure}

For the initial condition to simulate hydrodynamic equations we fit the data
of DEM at $t=20$ in the dimensionless unit. 
Each fitting curve is approximated by a polynomial of $y$ (Fig.\ref{fig:8-1}). 
The reason why we adopt the initial condition at $t=20$ instead of
$t=0$, we are interested in the slow evolution of hydrodynamic variables 
after the total kinetic energy is relaxed to be a constant.

\begin{figure}
\centering
\includegraphics[width=8cm]{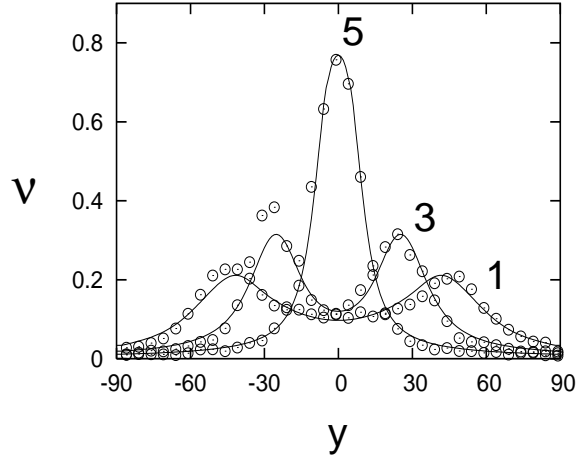}
\caption{The comparison of the data for the area fraction 
for $\bar\nu=0.121$ obtained by DEM (open circles) at $t=20$ (the label
 1), 60 (the label 3) and 380 (the label 5),
 and the result of
 hydrodynamic equations (\ref{hydro1})-(\ref{hydro3}) (solid lines).
}
\label{fig:8-2-3}
\end{figure}

\begin{figure}
\centering
\includegraphics[width=16cm]{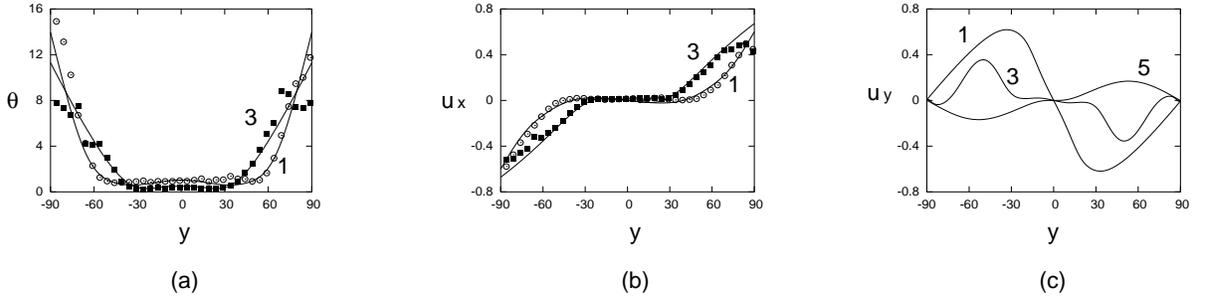}
\caption{The time evolution of the granular temperature (a), the velocity fields $u_x$ (b) and
 $u_y$ (c) in hydrodynamic simulations shown in solid lines, where the
 data in (a) and (b) are obtained by DEM.
The numbers
 1,3,5 in these figures correspond to results at $t=20$, 60 and 380, respectively. The DEM
 data with the solid squares and open circles correspond to the result
 at $t=20$ and 60, respectively. 
Note that comparison of the theory and DEM in the steady values of $\theta$
 and $u_x$ will be shown in the next section, while we do not include
 DEM data for $u_y$ because of the disagreement in the scale (see Fig.4).}
\label{fig:8-5-6}
\end{figure}

As shown in Figs.~\ref{fig:8-2-3} and \ref{fig:8-5-6} the results of the simulation of hydrodynamic
equations well agree with those of DEM. Once we rescale the time,
the evolution of hydrodynamic variables in the simulation of hydrodynamic equations is almost equivalent to that of DEM.
This agreement between hydrodynamic equations and DEM means that the
renormalization procedure by Yoon and Jenkins\cite{Yoon} gives us the
accurate results.  Amongst hydrodynamic variables the
$y-$component of the velocity field in the simulation of hydrodynamic
equations has much larger  than that of DEM though the profile
itself is similar with each other, but the other variables in hydrodynamic
simulation are almost the same as those in DEM (Fig.\ref{fig:8-5-6}). 

It should be noted that the transient dynamics of a granular shear flow
has been discussed by Babic\cite{babic97b} but his system is relaxed to
be an USF because of the small system size and inelasticity. On the other 
hand, ours will not reach USF, and the time evolution of hydrodynamic variables
contains a pattern formation.  

\subsection{Simulation of a simplified set of equations}

To understand the qualitative behavior of phase separations, we need to
reduce the degree of freedom of hydrodynamic equations. It is reasonable
to deduce the terms proportional to the bulk viscosity is not
important. In addition, the advection term $\Vect{u}\cdot\Vect{\nabla}
\Vect{u}$  in hydrodynamic equations may not play important roles because
we are interested in the slow dynamics in the domain growth.
The coupling between the spatial gradient and the terms proportional
to $1-e^2$ because the kinetic theory can be applied to cases for small
inelasticity.

Thus, we may reduce the set of hydrodynamic equations to be
\begin{eqnarray}
  \partial_t \nu &=& -\partial_y(\nu u_y),   \label{simple1} \\
  \partial_t u_x &=& \displaystyle \partial_y[\frac{\eta(\nu)}{2}\theta^{1/2}\partial_y u_x]  ,\label{simple2}\\
  \partial_t u_y &=& \partial_y[\eta(\nu)\theta^{1/2}\partial_y u_y - p(\nu)\theta]  ,\label{simple3}\\
  \partial_t \theta &=& -u_y\partial_y \theta - \nu^{-1}\eta(\nu)\theta^{1/2}(\partial_y u_x)^2
+2\nu^{-1}\eta(\nu)\theta^{1/2}(\partial_y u_y)^2 -
2\nu^{-1}\partial_y(\kappa(\nu)\theta^{1/2}\partial_y\theta) \nonumber\\
	&&-\sqrt{2\pi}(1-e^2)\nu g(\nu)\theta^{3/2} . \label{simple4}
\end{eqnarray}
For further simplification, we may assume that the temperature $\theta$ is a fast variable to
slave other slow variables. Thus, we may omit the time derivative in  eq.(\ref{simple4}), but such
simplification does not lead to a simplified treatment to solve hydrodynamic equations.
Although $u_y$ becomes zero in the steady state, it plays an important
role for the time evolution of density.  Thus, we believe that the set of equations (\ref{simple1})-(\ref{simple4})
is the simplest set of hydrodynamic equations to describe phase separations.

\begin{figure}[htbp]
\centering
\includegraphics[width=16cm]{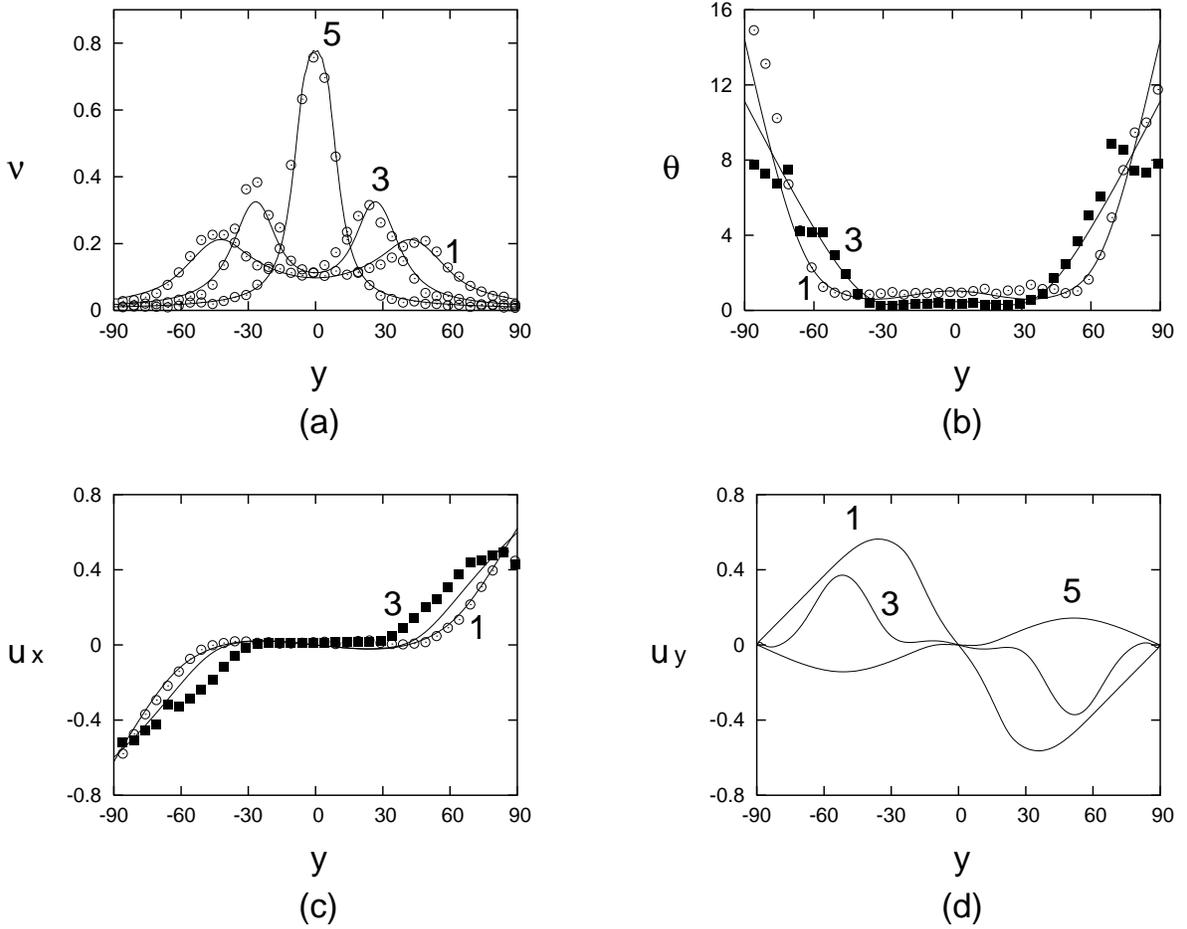}
\caption{The time evolution of the the area fraction (a) granular
 temperature (b), the velocity fields $u_x$ (c) and
 $u_y$ (d) in the simulation of a simplified model
 (\ref{simple4}) of hydrodynamic equations.
The numbers
 1,3,5 in these figures correspond to the data at 
$t=20, 60$ and 380, respectively.}
\label{fig:8-7-10}
\end{figure}

Figure~\ref{fig:8-7-10} shows the growth of hydrodynamic variables based on (\ref{simple1})-(\ref{simple4}).
Although the quantitative behavior is a little deviated from the result
of DEM or the full set of hydrodynamic equations
(\ref{hydro1})-(\ref{hydro3}) in particular for $u_x$, qualitative behavior of this simplified model is similar to those in
more accurate treatments. In the steady state both hydrodynamic models reduce to equivalent
results.

\section{Theoretical description of the steady state}

In the steady state, the hydrodynamic variables depend only on $y$. Thus, the variables are
\begin{equation}
 \nu=\nu(y),\hspace*{3em}u_x=u_x(y),\hspace*{3em}u_y=0, \hspace*{3em}\theta=\theta(y).
\end{equation}
It is obvious that any hydrodynamic variable $\Psi$ 
satisfies $D_t\Psi=\partial_t\Psi=0$. Thus, the equation of mass conservation is automatically satisfied.
The remain equations of motion become
\begin{eqnarray}
  0 &=& \displaystyle\frac{d}{dy}P_{xy} , \\
  0 &=& \displaystyle\frac{d}{dy} P_{yy} ,\\
  0 &=& \displaystyle P_{yx}\frac{d}{dy} u_x+\frac{d}{dy} q_y+\chi .
\label{steady}
\end{eqnarray}
Thus, the normal stress and the shear stress  are uniform 
\begin{equation}
p\equiv P_{yy}=const. \quad \tau\equiv P_{yx}=const.
\end{equation}
From the definition of the pressure tensor we obtain
\begin{eqnarray}\label{du/dy-theta}
 \tau &=& \displaystyle -\frac{\eta(\nu)}{2}\theta^{1/2}\frac{d u_x}{dy} , \\
 p &=& \displaystyle \frac{1}{2}\nu[1+(1+e)\nu g(\nu)]\theta .
\end{eqnarray}
Thus, we obtain the expressions for $\theta$ and $du_x/dy$ as functions of $p$, $\tau$ and 
$\nu$.
Substituting them into the last equation of (\ref{steady}) we obtain
\begin{equation}\label{2nd-eq}
 \frac{d}{d y}[F(\nu)\frac{d\nu}{dy}]=G(\nu) ,
\end{equation}
where 
\begin{eqnarray}
 F(\nu)&=&\frac{1}{\alpha(\nu)^{3/2}}[(\frac{1}{2}+r\frac{d}{d\nu}(\nu^2
  g))\frac{\kappa(\nu)}{\alpha(\nu)}-\lambda(\nu)] , \\  
 G(\nu)&=&\epsilon\frac{2\alpha(\nu)^{1/2}}{\eta(\nu)}-(1-e)\frac{\xi(\nu)}
 {\alpha(\nu)^{3/2}}
\end{eqnarray}
with $\epsilon=(\tau/p)^2$ and $\alpha(\nu)=2/(\nu[1+(1+e)\nu g(\nu)])$.

It is well established to solve the second order ordinary differential equation such as (\ref{2nd-eq}).
Introducing $H(\nu)$ as $dH(\nu)/d\nu=F(\nu)$ and the multiplying $dH/dy$ in both sides of 
(\ref{2nd-eq}), and thus integrate the equation from $y=0$ to $y$ we obtain
\begin{equation}
\frac{1}{2}\left(\frac{dH}{dy}\right)^2=\int_{\nu(0)}^{\nu(y)}d\nu F(\nu)G(\nu) ,
\end{equation}
where we use the symmetric condition $d\nu/dy=dH/dy=0$ at $y=0$.
\begin{eqnarray}
 \pm\int_{\nu(0)}^{\nu(y)} 
\frac{F(\nu)}{\sqrt{2\int_{\nu(0)}^{\nu}
F(\nu')G(\nu')d\nu'}}d\nu &=& y .
\label{steady-sol}
\end{eqnarray}
Thus, we obtain the equation of $y$ as the function of $\nu$.

To draw the actual profile of $\nu$, we start from a trial $\nu_1(0)$ to
integrate (\ref{steady-sol}) and calculate
$I_1=\int_{-\Delta/2}^{\Delta/2}\nu_1(y)dy$, where the suffix 1
represents the first trial function. Then we replace $\nu_1(0)$ by $\nu_2(0)$ to
reduce the deviation between $I_1$ and $\bar\nu$. We repeat this
relaxation procedure to obtain the converged result $I_M\to \bar\nu$
until $M$ th trial. Once we obtain $\nu$, we can determine $\theta$ and
$du_x/dy$ by  eq.(\ref{du/dy-theta}).

\begin{figure}
\centering
\includegraphics[width=16cm]{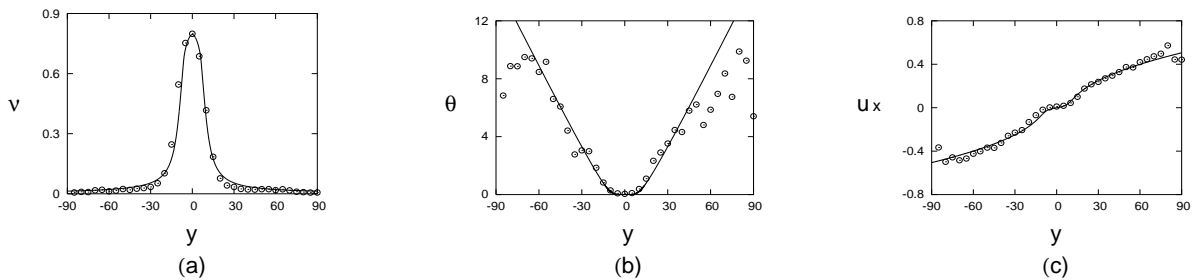}
\caption{The comparison between theory (solid lines) and DEM simulation
 without the tangential interaction (open circles) for area fraction (a),
 granular temperature (b) and
 $u_x$ (c). The mean area fraction is $\bar\nu=0.121$.}
\label{fig:6-1-3}
\end{figure}

\begin{figure}
\centering
\includegraphics[width=16cm]{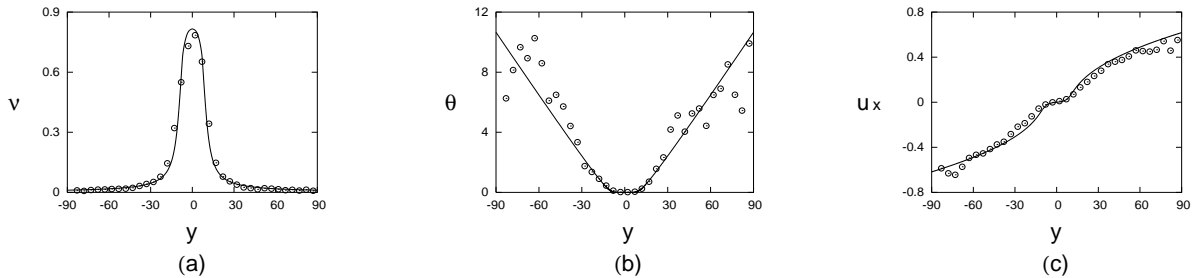}
\caption{The comparison between theory (solid lines) and DEM simulation
 with the tangential interaction (open circles) for area fraction (a),
 granular temperature (b) and
 $u_x$ (c). Here we use $e=0.798$  and the mean area fraction  $\bar\nu=0.121$.}
\label{fig:6-4-6}
\end{figure}

To compare $\nu$ and $du_x/dy$ with the result of DEM we use a 
fitting parameter $\epsilon=(\tau/p)^2$, while we need two fitting
parameters ($\tau=-0.0017$ and $p=0.1$ for non-rotational
cases and $\tau=-0.0017$ and $p=0.06$ for rotational cases)  
to determine $u_x$ and $\theta$. It should be noted that $p$ and $\tau$ are
determined by the boundary condition, but the boundary condition in
eqs.(\ref{8.4}) and (\ref{8.6}) with (\ref{8.7}) contains two
undetermined parameters. Thus, $p$ and $\tau$ cannot be determined
within our theory.
Figure \ref{fig:6-1-3} is  the comparison of our theoretical result with the result of DEM without the tangential
contact force in the interaction between particles for $\bar\nu=0.121$. The agreement between the theory and DEM is
good. Figure \ref{fig:6-4-6} is the comparison of our theoretical result with the result of DEM including
the tangential interaction taking into account the renormalization of
the restitution constant for $\bar\nu=0.121$.
We again obtain a fairly good agreement between the theory and the simulation. 



The reason why we can use the kinetic theory is that collisions between
particles are almost binary even in the dense cluster. Actually, we find 
that contacted particles are about 1.014 $\%$ of all particles at an
instance,  and only 
2.4 $\%$ of contacts are multi-body contacts among all
contacts for $\bar{\nu}=0.121$. 
Therefore, the kinetic theory can be used even in the dense
cluster in which particles are almost motionless. 

%

In principle, we can measure both the normal stress and the shear stress 
from the data of DEM. However, we only obtain the results with large
errors. It seems that there is a tendency to have too small $p$ in the
direct measurement, though observed $\tau$ is similar to the fitting value. 


Thus, we confirm that (i) the kinetic theory by Jenkins and
Richman\cite{JR} can reproduce the profiles of hydrodynamic variables 
to describe the
steady state of the granular fluid though the fitting values of the
stresses are included, 
and (ii) the renormalization scheme
proposed  by Yoon and Jenkins\cite{Yoon} is accurate.   
Although the setup of our simulation  seem to be  similar to  that
by Xu {\it et al}.\cite{Xu2}, our result for hydrodynamic variables 
 is much heterogeneous than that 
by Xu {\it et al}.  in the presence of a streamwise flow. 
In fact, our system is separated into two regions
which are a compact cluster and the very dilute region. 
The particles within the cluster is motionless, but the particles in the 
dilute region have large kinetic energy. On the other hand, the steady
state obtained by Xu {\it et al}. is similar to USF.
Thus, the result strongly depends on whether there is a stream flow.

\section{Dicussion}

Let us discuss our results in this section. To clarify the points we
discuss three important points; the linear stability analysis of USF, the
validity of the truncation at Navier-Stokes order and the effects of external
fields.

\subsection{The linear stability analysis}

Our analysis presented here suggests that USF of granular fluid is
unstable. To verify such suggestion, we need to discuss the linear
stability of USF. 

The stability of sheared granular flows has been discussed by many
researchers.\cite{Alam1,Alam2,Alam3,Linear1,Linear2,Linear3,Linear4,kumaran2,garzo06}
Many of them assume that the granular fluid is confined in an 
infinitely large system, but Alam {\it et al.}\cite{Alam1} discuss
the bifurcation of the steady solution as a function of the system size.
We should note that the stability of granular flows on an inclined slope 
has also been discussed by some researchers.\cite{FP,Mitarai1}

Here, we have also checked the linear stability of the uniform sheared
state assuming  Lees-Edwards sheared boundary condition. Since the result is
almost equivalent to that by Alam and Nott\cite{Alam3}, we
only summarize the outline of our linear stability analysis.
Our result indicates  that  USF 
is always unstable for enough large systems, but can be stable if the
system size is enough small near $e=1$. This result is consistent with
the results by Babic\cite{babic97b} and Popken and
Cleary\cite{Popken}. The details of our linear stability analysis will
be reported elsewhere.

As indicated in Section I,  Bagnold's scaling can be used if the
system is uniform and the heat conduction is negligible. However, the
granular gas  under the plane shear is not the case that we can assume
Bagnold's scaling. The heat conduction plays an important role and USF
cannot be maintained even when we adopt Lees-Edwards boundary condition.

\subsection{The validity of the approximation at Navier-Stokes order}

In this paper, we use the set of hydrodynamic equations at Navier-Stokes 
order. It should be noted that the set of hydrodynamic equations at
Navier-Stokes order does not mean Newtonian. Actually, our granular system
exhibits non-Newtonian behavior {\it i.e.} non-uniform velocity gradient
as in Figs.9 (c) and 10(c) in spite of the uniform  shear stress in the
steady state.

From our analysis, the effects from
the contraction of phase volume in collisions in  the inelastic
Boltzmann-Enskog equation are also small. The effects of tangential
contact force and the rotation of particles are also not important in
the bulk behavior of hydrodynamics.
Therefore, the kinetic theory
by Jenkins and Richman\cite{JR} gives us sufficient accurate results to
describe the hydrodynamics.

Santos {\it et al.}\cite{santos04} and 
Tij {\it et al}.\cite{tij} indicate that the transport coefficients in
Couette flow depend on the dimensionless shear rate. However, in our
system the shear rate only determines the time scale and thus 
the qualitative  behavior should not depend on the shear rate. 
There is room for discussion on the role of the shear rate in granular
gases as an open problem.

Some persons suggest the relevant role of terms at Burnett or super-Burnett
order terms\cite{kumaran2,Sela96,Sela98}, but our analysis indicate that 
the contribution from higher oder terms should be small.  It is known
that hydrodynamic equations at Burnett order derived from
the classical gas kinetics is unstable for perturbation, {\it i.e.} some
solutions will  blow up when a perturbation is applied to the
system.\cite{bobylev,struchtrup}  Therefore, we may need to be careful
to use hydrodynamic equations at Burnett order or super-Burnett order to 
describe granular fluids.

\subsection{The effects of external fields }

Our system does not contain the external force except for the driving
force of the boundary plates. In physical situations, it is difficult to 
remove the effect of gravity and collisions among particles may not be
regarded as binary. As indicated in Introduction, recent papers and
cited therein\cite{GDR,cruz}
of sheared granular materials under a constant pressure produce a state
of 'granular liquid' in which particles are multiple contacts with each
other.  In these cases, the square root 
of stiffness of grains becomes a characteristic frequency. Thus, the
behavior should depend on the shear rate. 

If we are interested in jamming transition or related
phenomena of granular materials, we need to apply an external field to
the system. Such subject will be important even in practical sense.

\section{Conclusion}

In this paper, we have confirmed the  validity of
hydrodynamic equations derived from the kinetic theory by Jenkins and
Richman.\cite{JR} We also confirm the  relevancy of the
renormalization method of the restitution constant by Yoon and
Jenkins.\cite{Yoon} This result may be surprised because the system
includes a dense cluster whose packing fraction is close to the maximum
value, and the particles in the cluster are almost motionless. 
Since USF is unstable, 
we cannot use Bagnold's
scaling to characterize the granular fluid in this case.

\vspace*{0.5cm}

We thank N. Mitarai and T. Hatano for fruitful discussion and their
variable comments. HH also
thanks J. T. Jenkins for his useful comments. 
This study is partially supported by the Grants-in-Aid of Japan Space
Forum, and
Ministry of Education, Culture, Sports, Science and Technology (MEXT), 
Japan (Grant No. 15540393 and No. 18540371) and the Grant-in-Aid for the 21st century COE 
"Center for Diversity and Universality in Physics" from MEXT, Japan.

\appendix
\section{The derivation of Walton's $\beta_0$}

In this Appendix, we demonstrate how to derive $\beta_0$ in Walton's
expression in eq.(\ref{walton})\cite{walton} for the tangential restitution coefficient.  
Although the theory by Maw {\it et al.}\cite{maw1,maw2,stronge,kuninaka}
has been used to evaluate $\beta$, their expression is complicated and
has an implicit form. Therefore, an explicit expression for Walton's
$\beta_0$ and the critical angle $\gamma_c$ is useful. 

Let us consider a collision between identical disks. Following the
notation in Section II (without suffices $i$ and $j$ for colliding
particles), the equation of motion for the tangential
direction is described by
\begin{equation}\label{3-14}
\ddot{w}_t+2(\eta_t \dot{w}_t+k_t w_t)=0,
\end{equation}
when there is no slip during the collision.
The factor 2 appears because the reduced mass is the half of mass of
each particle. The solution of eq.(\ref{3-14}) is easily obtained as
\begin{equation}
w_t=\frac{w_t(0)}{b}e^{-\eta_t t}\sin (bt), \quad 
\dot{w}_t(t)=w_t(0)e^{-\eta_t t}(\cos(b t)-\frac{\eta_t}{b}\sin(bt))
\end{equation}
for $\eta_t^2<2k_t$, where  $b=\sqrt{2k_t-\eta_t^2}$.

\begin{figure}
\centering
\includegraphics[width=10cm]{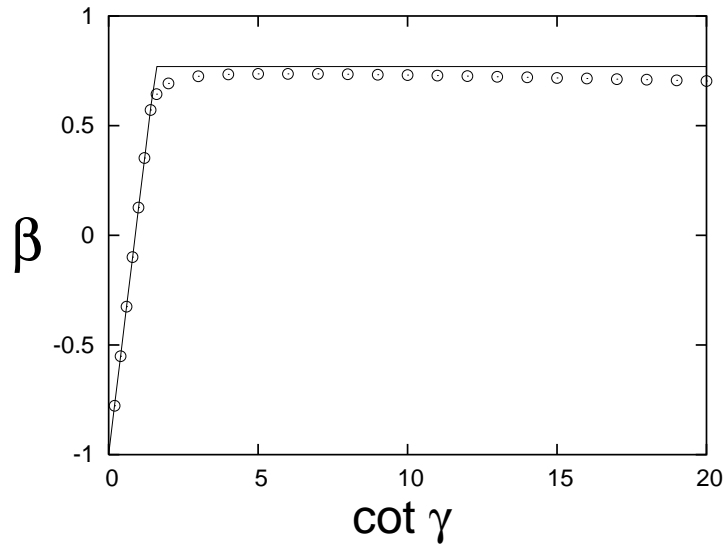}
\caption{The comparison of eq.(\ref{walton}) with (\ref{3-27}) (solid line) and the
 data obtained from DEM with $\mu=0.2$ (open circles). }
\label{fig:beta}
\end{figure}

Since we choose large $k_t$ and small $\eta_t$, the relation
$\eta_t^2<2k_t$ should be satisfied. Similarly, the equation of motion
in the normal direction is also described by an equation for a dumped
oscillation. Thus, the duration time $t_d$ at which $w_n=0$  is satisfied
and the normal restitution constant are respectively given by
\begin{equation}
t_d=\frac{2\pi}{\sqrt{2k_n-\eta_n^2}}, \quad \bar{e}=\exp[-\frac{\pi \eta_n}{\sqrt{2k_n^2-\eta_n^2}}] .
\end{equation}
On the other hand, the restitution constant $\beta_0$ for the tangential
contact is thus given by
\begin{eqnarray}\label{3-27} 
\beta_0&=& -\frac{\dot{w}_t(t_d)}{\dot{w}_t(0)} \nonumber \\
&=& \exp(-\frac{\pi A\eta_t}{\eta_n\sqrt{2R-1}})
[\frac{1}{\sqrt{2Q/A-1}}\sin(\frac{\pi A\eta_t\sqrt{2Q/A-1}}{\eta_n\sqrt{2R-1}})\nonumber\\
&&-\cos(\frac{\pi A\eta_t\sqrt{2Q/A-1}}{\eta_n\sqrt{2R-1}})] ,
\end{eqnarray}
where 
\begin{eqnarray}
	A=1+\frac{m\sigma^2}{4I} ,\hspace*{2em} 
	R=\frac{\pi k_n}{4\eta_n^2} , \hspace*{2em}
	Q=\frac{\pi k_t}{4\eta_t^2} . 
\end{eqnarray}
If we substitute the values $k_n=3.0\times 10^3$, $k_t=k_n/4$,
$\eta_n=3.0$ and $\eta_t=\eta_n/2$ used in DEM simulation, we obtain
$\beta_0\simeq 0.769235$. The comparison between the theory
(\ref{walton}) with (\ref{3-27}) 
and DEM is shown in Fig.~\ref{fig:beta}. Without the introduction of any 
fitting parameters, agreement between the theory and DEM is fairly
good. Here the critical angle
$\cot\gamma_c=(1+\beta_0)\mu(1+\bar{e})\simeq 1.56734 $.

\end{document}